\documentclass[12pt, onecolumn, draftcls]{IEEEtran}
\usepackage{amsfonts}
\usepackage{amssymb}
\usepackage{cite}
\usepackage{graphicx}
\usepackage{psfrag}
\usepackage{subfigure}
\usepackage{amsmath}
\usepackage{array}
\usepackage{fancyhdr}

\newcommand{\whp}{\textit{whp} }

\newtheorem{Lemma}{Lemma}
\newtheorem{Theorem}{Theorem}

\newtheorem{Corollary}{Corollary}
\newtheorem{Scheme}{Scheme}
\newtheorem{Baseline}{Baseline}

\begin{document}

\title{\huge A New Scaling Law on Throughput and Delay Performance of
Wireless Mobile Relay Networks over Parallel Fading Channels}

\author{Rui Wang,  Vincent K. N. Lau, and Kaibin Huang\\
Department of Electronic \& Computer Engineering\\
Hong Kong University of Science \& Technology\\
 Clear Water Bay, Hong Kong\\
Email:  \{wray, eeknlau\}@ust.hk, khuang@ieee.org\\
\thanks{This work was supported by the Research Grants Council of the
Hong Kong Government through the grant RGC 615407.} }

\maketitle

\begin{abstract}
In this paper, utilizing the relay buffers, we propose an opportunistic
decode-wait-and-forward relay scheme for a point-to-point communication system
with a half-duplexing relay network to better exploit the time diversity and
relay mobility. For instance, we analyze the asymptotic throughput-delay
tradeoffs in a dense relay network for two scenarios: (1) fixed relays
with \textit{microscopic fading } channels (multipath channels), and
(2) mobile relays with \textit{macroscopic fading} channels (path
loss). In the first scenario, the proposed scheme can better exploit
the \textit{multi-relay diversity} in the sense that with $K$ fixed
relays and a cost of $\mathcal{O}(K)$ average end-to-end packet
delay, it could
achieve the same optimal asymptotic average throughput as the
existing designs (such as regular decode-and-forward relay schemes)
with $K^2$ fixed relays. In the second scenario, the proposed scheme
achieves the maximum throughput of $\Theta(\log K)$ at a cost of
$\mathcal{O}(K/q)$
average end-to-end packet delay, where $0<q\leq \frac{1}{2}$ measures
the speed of relays'
mobility. This system throughput is unattainable for the
existing designs with
low relay mobility, the proposed relay scheme can exploit the relays'
mobility more
efficiently.
\end{abstract}

\section{Introduction}\label{sec:intro}

In wireless communication networks, cooperative relaying not only
extends the coverage but also contributes the \textit{spatial
diversity}. As a result, cooperative relay is one of the core
technology components in the next generation wireless systems such
as IEEE 802.16m and LTE-A. Extensive research is being carried out
on the theory and algorithm design of cooperative relay and there
are still a lot of open issues associated with the capacity and
the design of relay networks. For instance, most of the existing works on relay
protocol designs have ignored and failed to exploit the buffer dynamics in the
relay. As such, how could the
throughput of relay networks benefit from the mobility of relays
is still not clearly known and we shall try to shed some lights on
this aspect in this paper. Specifically, utilizing buffers in the relays, an
opportunistic
decode-wait-and-forward (ODWF) relay scheme is proposed to
exploit the mobility of relays so as to improve the system throughput
of the mobile relay networks. Moreover, the asymptotic tradeoff of
the end-to-end packet delay versus the end-to-end throughput of
the ODWF scheme is analyzed to obtain the design insights.

\subsection{Related Work}
The initial study of relay channels can be traced back to 1970s
\cite{Meulen:68,Meulen:71,Meulen:77}. The capacity of the relay
channel consisting of single source, destination and relay station
was derived by Cover and El Gamal in \cite{Cover:79}. Since then,
different types of relay channels have been investigated from the
information theoretic perspective, including the relay-assisted
point-to-point communication
\cite{GamalAE:82,ZhangZhen:88,LiangliangXie:05}, relay broadcasting
channels in
\cite{Reznik:05,Kramer:05,YingbinLiang:07}, multi-source
multi-destination relay channels
\cite{Kramer:05,HostMadsen:03,HostMadsen:06}. Recent research
mainly focuses on two topologies of wireless relay (or multi-hop)
networks, namely \textit{ad hoc} networks and relay-assisted
point-to-point communications. In wireless \textit{ad hoc}
networks, any pair of nodes are allowed to communicate; in
relay-assisted point-to-point communications, there exists only
one pair of source and destination while other nodes serve as
relay stations. The scaling law of the wireless \textit{ad hoc}
network capacity is first studied by Gupta and Kumar in
\cite{Gupta:00}. Specifically, it is shown that each node can
achieve the throughput of the order
$\mathcal{O}(\frac{1}{\sqrt{K\log K}})$ \footnote{Throughout this
paper, we shall use the following asymptotic notations to
demonstrate the orderwise relationship. $f(K)=\mathcal{O}(g(K))$
if $f(K)\leq cg(K)$ for some constant $c$ and sufficiently large
$K$. $f(K)=\Theta(g(K))$ if $f(K)=\mathcal{O}(g(K))$ and
$g(K)=\mathcal{O}(f(K))$.} when $K$ fixed nodes are randomly
distributed over a unit area. The same throughput bound is also
obtained in \cite{GamalAE:04,Toumpis:04,Kulkarni:04} by other
approaches. Later, in \cite{Franceschetti:07} the throughput of
the order $\mathcal{O}(\frac{1}{\sqrt{K}})$ is shown to be
achievable, and in \cite{Aeron:07} the throughput order is
enhanced to $\mathcal{O}(\frac{1}{K^{\frac{1}{3}}})$. These
results imply that the throughput of each node converges to zero
when the number of nodes increases. Nevertheless, it's found in
\cite{Ozgur:07} that the per-node throughput can arbitrarily close
to constant by hierarchical cooperation, and in
\cite{Grossglauser:02} that the constant per-node throughput is
achievable by exploiting the node mobility. In addition, in
\cite{GamalAE:04,Toumpis:04} the throughput-delay tradeoff of
wireless \textit{ad hoc} networks is studied as the gain on
network throughput is obtained at the expense of packet delivery
delay.

The design of point-to-point communication via relay networks
differs from that of wireless \textit{ad hoc} networks in that
there is only one data stream in the system and the system
objective is to maximize single-link throughput, rather than the
throughput of the weakest source-destination pair (in wireless
\textit{ad hoc} networks). Hence, it's expected that the
relay-assisted point-to-point communications could achieve higher
per-node throughput than \textit{ad hoc} networks. In
\cite{Gastpar:02b,Gastpar:05}, it's shown that the
source-destination throughput can scale as $\log K$ when all the
relays amplify and forward the received packet to the destination
cooperatively with $\Theta(K)$ total transmission power. This
relay strategy was extended to the case of multiple antennas in
\cite{Bolcskei:06}, where the throughput can scale as
$\frac{M}{2}\log K$ with $M$ antennas at both the source and the
destination. The throughput scaling with $\Theta(\log K)$ as
derived in prior works can be interpreted as follows: since all
the relay nodes should amplify and forward the received packets,
the total transmit power of the relay network is $\Theta(K)$, then
by Shannon's capacity theory the throughput is $\Theta(\log K)$.

\subsection{Contributions}

In this paper, we shall extend the existing knowledge on the
point-to-point communications with relay networks in the following
aspects.
\begin{itemize}
\item By exploiting the mobility of relay nodes through our
proposed ODWF scheme, the asymptotical throughput $\Theta(\log K)$
is also achievable even with $\Theta(1)$ total transmit power and
decode-and-forward relays. Moreover, compared with the constant
per-node throughput (achieved by exploiting the relays' mobility)
in \cite{Grossglauser:02}, our work shows that if there is only
one data stream in the system, the source can transmit with a much
higher date rate with the assistance of other relays.

\item Even if the relays are fixed, the ODWF scheme can better
exploit the \textit{multi-relay diversity}, therefore, achieving
higher throughput than the existing designs.
\end{itemize}

\begin{figure}
 \begin{center}
  \resizebox{15cm}{!}{\includegraphics{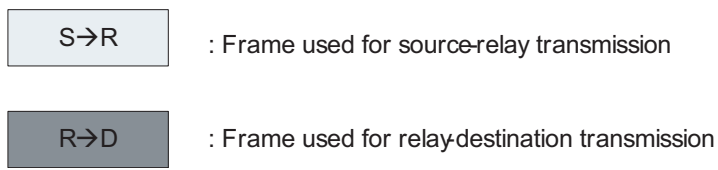}}
 \end{center}
    \caption{Frame sequences for regular decode-and-forward design and opportunistic decode-wait-and-forward design.}
    \label{fig:timing}
\end{figure}

Specifically, we propose an opportunistic decode-wait-and-forward
(ODWF) relay scheme which utilizes relay buffers to exploit the
mobility of relays in a point-to-point communication link assisted
by $K$ mobile relays with $N$ parallel fading channels (e.g. OFDM
systems). In this scheme, only one of the relays is selected at
any time to forward packets to the destination. The key feature of
the proposed scheme is that the phase I (source to relay) and
phase II (relay to destination) transmissions are scheduled based
on the instantaneous channel states rather than having phase I and
phase II as inseparable atomic actions (phase II always follows
immediately from phase I) as in the existing designs, which is
illustrated in Fig.\ref{fig:timing}. By exploiting the mobility of
relays in the phase I and phase II scheduling, we could achieve a
better throughput scaling law. Moreover, we also examine the
performance of the ODWF scheme in fixed relay networks, and it's
shown that the ODWF scheme could better exploit the
\textit{multi-relay diversity} than the existing designs even
without relay mobility. In the proposed ODWF scheme, packets may
have to be buffered at the relay nodes when phase II transmission
cannot be scheduled immediately. As a result, we shall quantify
the scaling law of the throughput-delay tradeoff in two scenarios,
namely the {\em fixed relays with microscopic fading} (multipath
fading) as well as the {\em mobile relays with macroscopic fading}
(path loss). As comparisons, we also derive the throughput of the
existing relay schemes (i.e. regular decode-and-forward schemes)
in both scenarios as a performance baseline. In the scenario of
fixed relays with microscopic fading, the proposed ODWF scheme
exploits the time diversity of the multipath channels by
scheduling the phase I and phase II transmissions adaptively. For
this scenario, the proposed scheme can achieve the following
performance:
\begin{itemize}
\item The achievable average system throughput is
$\frac{N}{2}\log_2 \left[1+p\ln K\right]$, which is higher than
$\frac{N}{2}\log_2 \left[1+p\ln \sqrt{K}\right]$ achieved by the
existing regular decode-and-forward scheme for $K$ fixed relays,
where $p$ is transmit power and $N$ is number of subcarriers.

\item The average system throughput $\frac{N}{2}\log_2
\left[1+p\ln \beta\right]$ can be achieved at a cost of
$\frac{\beta^2}{K}$ average
end-to-end packet delay, where $\beta \sim \mathcal{O}(K)$.
\end{itemize}
On the other hand, in the scenario of mobile relays with
macroscopic fading, the ODWF scheme exploits the relay mobility by
scheduling the phase I and phase II transmissions and it can
achieve the following performance:
\begin{itemize}
\item The proposed scheme achieves a higher throughput scaling of
$\frac{N\alpha}{4} \log_2K$ compared with the regular
decode-and-forward schemes. The throughput gain over the regular
schemes is $\mathbf{\Theta}\left(\frac{\log_2
K}{qK^{\frac{1}{M-1}}}\right)$ when the relays' velocity $q$ is
small (e.g. $q=\frac{1}{K}$), where $\alpha$ is the pathloss
exponent, $N$ is the number of subcarriers, $M>1$ is an integer
related to the relays' mobility model.

\item The average system throughput $\frac{N}{2}\log_2
\beta$ can be achieved at a cost
of
$\mathcal{O}(\max\{\frac{1}{q},\frac{\beta^{\frac{4}{\alpha}}}{Kq}\}
)$ average end-to-end packet delay, where $\beta \sim
\mathcal{O}(K^{\frac{\alpha}{2}})$.

\end{itemize}

The remaining of this paper is organized as follows. In Section
\ref{sec:model}, the system models for both fixed relays and
mobile relays scenarios are introduced. In Section
\ref{sec:rlt-micro}, the ODWF scheme for fixed relays scenario is
described, followed by the throughput-delay tradeoff analysis. In
Section \ref{sec:rlt-macro}, the ODWF scheme for mobile relays
scenario as well as the related throughput-delay tradeoff analysis
is presented, followed by the conclusions in Section
\ref{sec:con}.

\section{System Model}
\label{sec:model}

In this section, we shall describe the system models for both
fixed relays and mobile relays scenarios, and define the
performance metrics.

\subsection{System Model for Fixed Relays with Microscopic Fading}

\begin{figure}
 \begin{center}
  \resizebox{15cm}{!}{\includegraphics{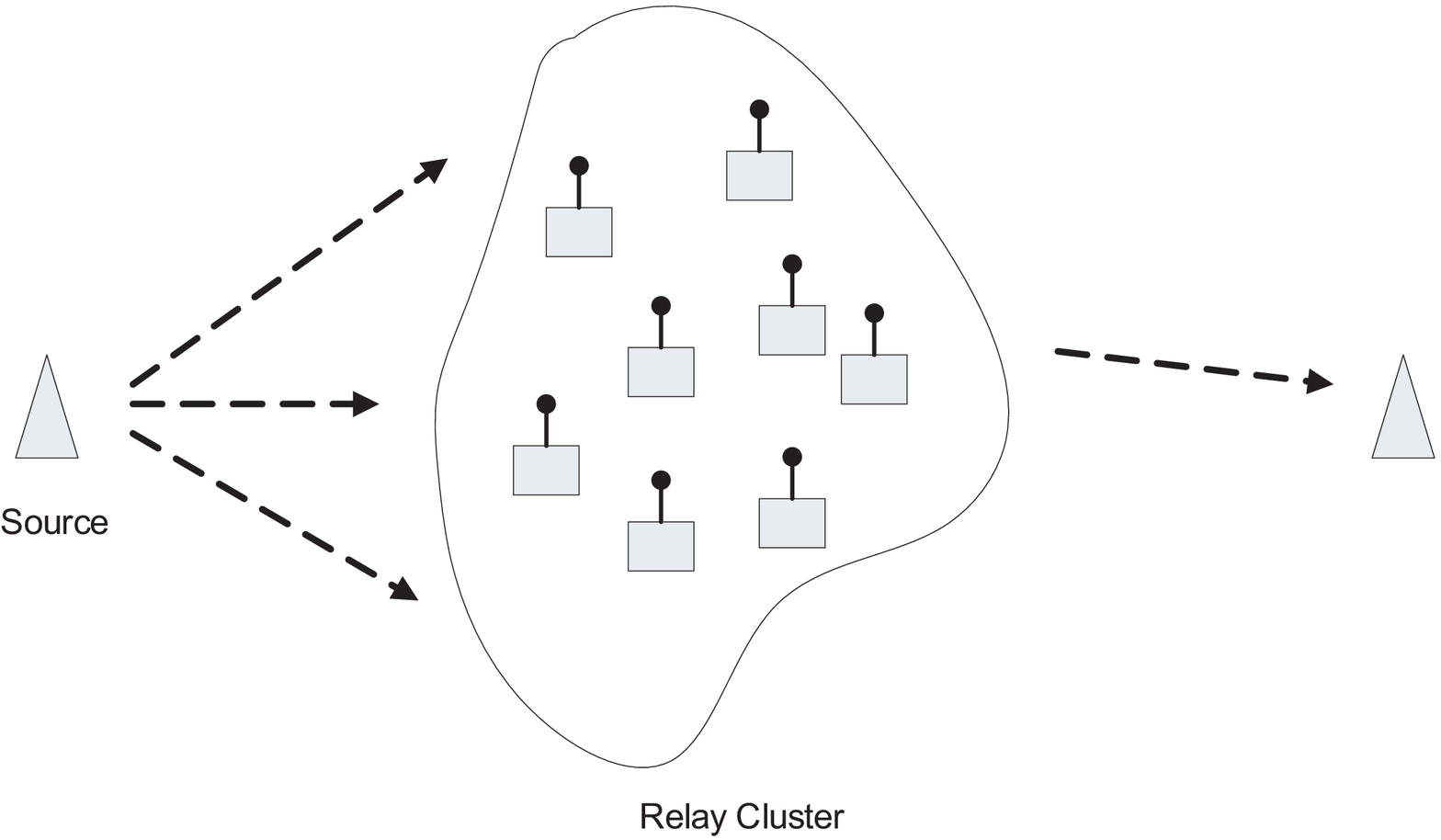}}
 \end{center}
    \caption{System model of the microscopic fading scenario}
    \label{fig:fix}
\end{figure}

\begin{figure}
 \begin{center}
  \resizebox{15cm}{!}{\includegraphics{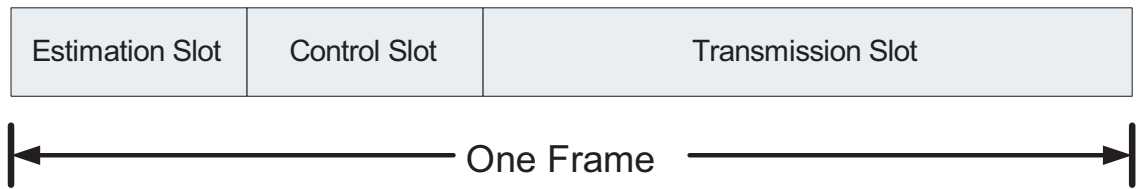}}
 \end{center}
    \caption{The illustration of frame structure, where the length of
transmission slot is much larger than that of the estimation slot
and control slot.}
    \label{fig:frame}
\end{figure}

In the scenario of fixed relays, a cluster of $K$ fixed relay
nodes lies between the source and the destination, as shown in
Fig. \ref{fig:fix}. The path loss of all source-relay and
relay-destination links are assumed identical. The OFDM technology
is used for converting the frequency selective channel into $N$
parallel flat fading channels (subcarriers). There is no direct
link between the source and destination due to the long distance
and hence the source could only rely on the relay network for
forwarding its packets to the destination (as in
\cite{Gastpar:05}). The relay strategy we consider is
decode-and-forward (DaF) \cite{Cover:79}, and all the relays are
half-duplexing. Moreover, at most one relay is selected at any
time on each subcarrier. To facilitate the relay scheduling,
transmission is partitioned into frames. As illustrated in Fig.
\ref{fig:frame}, each frame is further
divided into three types of slots defined as follows:
\begin{itemize}
\item {\bf Channel Estimation Slot} is used by relays for
estimating the channel gains with the source and destination.

\item {\bf Control Slot} is used by relays for distributive
control signaling of the ODWF scheme. The details is given in the
scheme description.

\item {\bf Transmission Slot} is used for data transmission.
\end{itemize}

\subsubsection*{Physical Layer Model}

We shall model the physical layer performance by means of mutual information.
Such an approach decouples the implementation details (coding and modulation) of
the PHY layer and moreover, it has been demonstrated that strong coding such as
LDPC code can achieve the Shannon's capacity to within $0.05$dB with reasonable
block size (e.g. 2k bytes) and target PER of $10^{-3}$.  The maximum mutual
information between
the source and the $j$th relay on the $n$th subcarrier is given as
\begin{equation}
C^{s,j}_n = \log_2 (1+p|H^{s,j}_n|^2), \nonumber\\
\end{equation}
and that between the $j$th relay and the destination on the $n$th
subcarrier is given as
\begin{equation}
C^{j,d}_n = \log_2 (1+p|H^{j,d}_n|^2), \nonumber\\
\end{equation}
where $p$ is the transmit power on each subcarrier, $H^{s,j}_n
\sim \mathcal{CN}(0,1)$ and $H^{j,d}_n \sim \mathcal{CN}(0,1)$ are
the microscopic fading gains (due to Rayleigh fading), and the
effect of path loss is neglected as it's identical for all links.
All the packets are transmitted at data rate $r$. The $j$th relay
could correctly decode the packets received over the $n$th
subcarrier from the source only if  $r\leq C^{s,j}_n$, and the
destination could correctly decode the packets received over the
$n$th subcarrier from the $j$th relay only if $r\leq C^{j,d}_n$.
For convenience, we shall call a link as a {\it connected link} if
its maximum mutual information is larger than $r$, and otherwise a
{\it broken link}.

\subsection{System Model for Mobile Relays with Macroscopic Fading}

In the scenario of mobile relays, we also consider a
point-to-point communication system with $K$ relays and $N$
subcarriers. However, in this scenario, the relays are moving and
the frame duration (the length of channel codeword) is much larger
than that in the fixed relay scenario in order to capture the
relays' mobility. Therefore, we highlight in the following the
differences of the system model for mobile relay networks from
that for fixed relay networks discussed in the preceding part.

\subsubsection*{Relay Mobility Model}

\begin{figure}
 \begin{center}
  \resizebox{15cm}{!}{\includegraphics{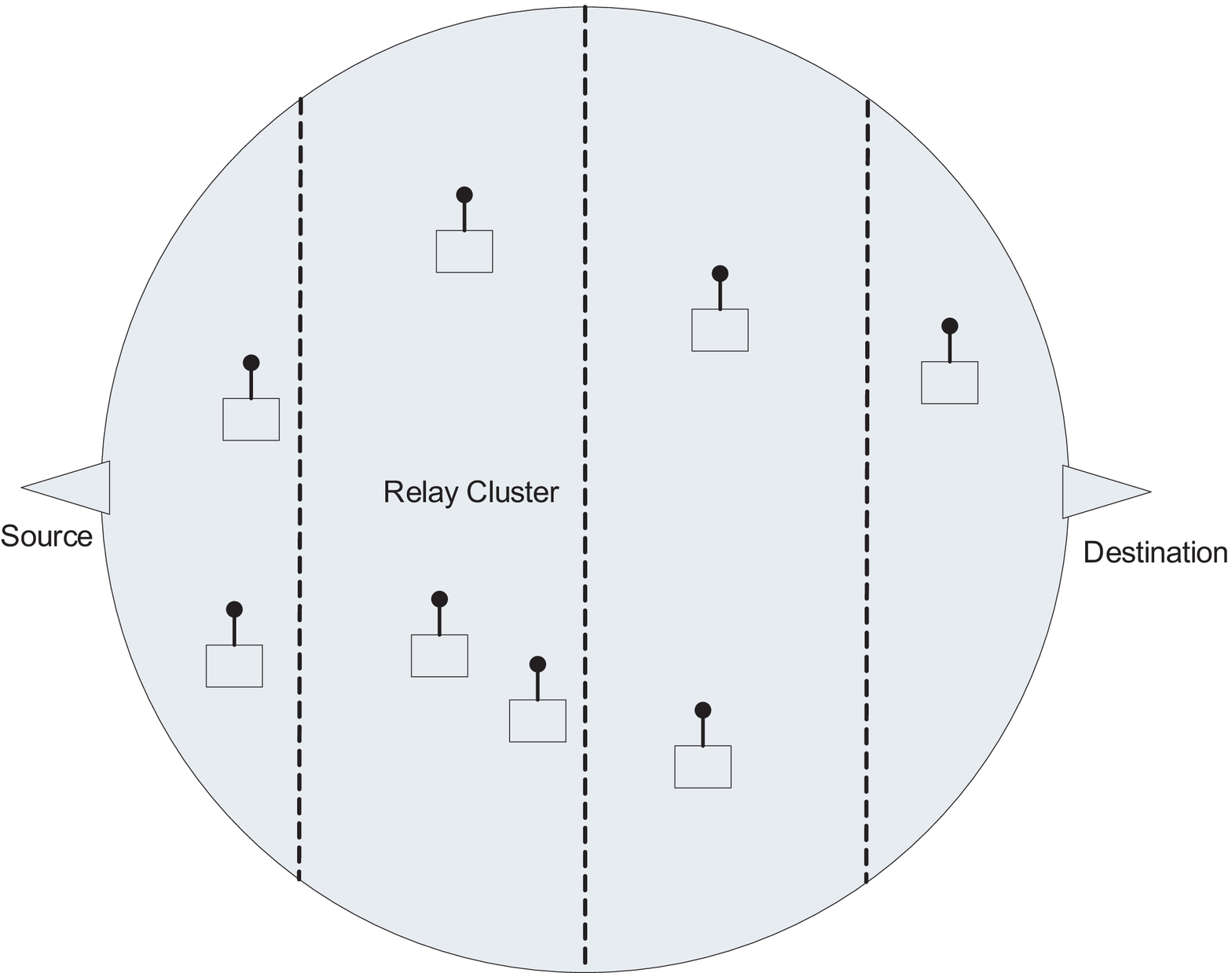}}
 \end{center}
    \caption{System model of the macroscopic fading scenario}
    \label{fig:mobile}
\end{figure}

\begin{figure}
 \begin{center}
  \resizebox{10cm}{!}{\includegraphics{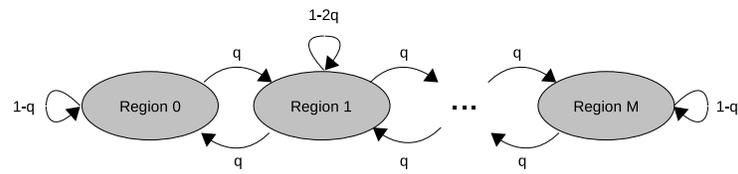}}
 \end{center}
    \caption{The random walk of each relay.}
    \label{fig:mchain}
\end{figure}

Following \cite{Gupta:00,Gastpar:05}, we assume that the $K$
relays are distributed on a disk with radius $R$ as illustrated in
Fig. \ref{fig:mobile}. The source and the destination are fixed at
two ends of a diameter, and the disk is divided horizontally into M
equal-area regions along the source-destination diameter. These
regions are denoted as region 1, region 2, ..., and region M,
from the source to the destination. As illustrated in Fig.
\ref{fig:mchain}, the movement
of each relay is modeled as a random walk (Markov chain) on these
regions:
\begin{itemize}
\item At the beginning, each relay is uniformly distributed on the
disk. Movements of relays can only occur in discrete frame with time index $t$.

\item Let $X_k(t)$ denote the region index of the $k$th relay in the
$t$th frame,
$\{X_k(t)|t=1,...+\infty\}$ is a Markov chain with the following
transition matrix
\begin{equation}
Q_{i,j}=\left\{ \begin{array}{ll}
q & j=i+1 \ \mbox{and} \ i=1,...,M-1\\
q & j=i-1 \ \mbox{and} \ i=2,...,M\\
1-2q & j=i=2,...,M-1\\
1-q & j=i=1 \mbox{ or } M\\
0 & \mbox{otherwise}
\end{array} \right.
\end{equation}
where $Q_{i,j}$ denotes the transition probability that one relay
jumps into the $j$th region in the next frame, providing that it
is in the $i$th region currently.

\item When one relay moves into a region, its actual location in
this region is uniformly distributed.
\end{itemize}
\textit{Remarks:} The region transition probability $q$ measures
how likely one relay will move into another region in the next
frame, and therefore, it is related to the average speed of the
relays.

\subsubsection*{Physical Layer Model}

In order to capture the dynamics of macroscopic fading (relays'
mobility), we assume the frame duration is much larger than the
coherent time of the
microscopic fading. As a result, the transmitter can deliver
packets with the ergodic capacity (averaged over multipath fading)
on every subcarrier in every frame. Since for one
transmitter-receiver pair the ergodic capacity of different
subcarriers is identical, it's of less interests to consider
different scheduling on different subcarriers. Hence, we assume
each transmitter (source or relay) will use the whole spectrum
when it's transmitting packets, and each packet is jointly encoded
across all subcarriers. Let $p$ be the transmit power on each
subcarrier, the ergodic capacity between the source and the $j$th
relay is given by
\begin{equation}
C^{s,j} = \mathbf{E}\left[\sum_{n=1}^N \log_2
(1+p\frac{|H^{s,j}_{n}|^2}{d_{s,j}^{\alpha}})\right]
\label{eqn:cp-sj} ,
\end{equation}
where $\mathbf{E}[\cdot]$ is the expectation over the random
Rayleigh fading gain $H^{s,j}_{n}$, $d_{s,j}$ is the distance
between the source and the $j$th relay, $\alpha$ is the path loss
exponent. Furthermore, the ergodic capacity between the $j$th
relay and the destination is given by
\begin{equation}
C^{j,d} = \mathbf{E}\left[\sum_{n=1}^N \log_2
(1+p\frac{|H^{j,d}_{n}|^2}{d_{j,d}^{\alpha}})\right]\label{eqn:cp-jd}
,
\end{equation}
where $d_{j,d}$ is the distance between the destination and the
$j$th relay, and $\mathbf{E}[\cdot]$ is the expectation over the
random Rayleigh fading gain $H^{j,d}_{n}$. Therefore, one link is
\textit{connected} when the packet data rate $r$ is less than or
equal to the ergodic link capacity defined in (\ref{eqn:cp-sj}) or
(\ref{eqn:cp-jd}), and \textit{broken} otherwise.

\subsection{Performance Measures}\label{sub:measure}

As explained in Section \ref{sec:intro}, packets may have to be
buffered in the relays in the proposed ODWF scheme. As a result,
we consider both the average system throughput and the average
end-to-end packet delay as our system performance measures. They
are defined rigorously below.

\textit{Average system throughput:} Let $T_1$ be the number of
information bits successfully received by the destination in a
particular frame after the system's running for sufficiently long
time. Moreover, let $T_2,T_3,...,T_S$ be the number of information
bits successfully received by the destination in the following
$S-1$ frames. Hence, the average system throughput $T$ is defined
as
\begin{equation}
T = \lim_{S\rightarrow +\infty}\frac{\sum_{i=1}^S
T_i}{S}.\label{def:avg-tp}
\end{equation}

\textit{Average end-to-end packet delay:} Suppose the system has
been running for sufficiently long time, let $I^t_{1}$ and
$I^r_{1}$ be the indices of the frames in which certain packet is
transmitted from the source and successfully received by the
destination respectively. Also, let
$\{I^t_{2},I^r_{2}\},\{I^t_{3},I^r_{3}\},...,\{I^t_{S},I^r_{S}\}$
denote the frame indices of the following $S-1$ packets, the
average end-to-end packet delay is defined as
\begin{equation}
D = \lim_{S\rightarrow +\infty}\frac{\sum_{i=1}^S
I^t_{i}-I^r_{i}}{S}.\label{def:avg-delay}
\end{equation}
Notice that this packet delay measures the average time one packet
stays in the relay's buffer. We do not consider queueing delay at
the source and delegate it to future work.

Moreover, due to the randomness of the system model, there is no
deterministic relationship on the average system throughput and
the average end-to-end packet delay. Therefore, we shall use the
notation $\doteq$ to denote the equality with probability
$1-\frac{1}{K}$ (or with high probability, \whp) as in
\cite{GamalAE:04}. Clearly, when the number of relays $K$ tends to
infinity, the probability the equality holds will tend to $1$.

\section{Opportunistic Decode-Wait-and-Forward Scheme for Fixed
Relays with Microscopic
Fading Channels}\label{sec:rlt-micro}

In this section, we first describe the ODWF relay scheme for the
scenario of fixed relays and then, analyze its performance. Utilizing the
buffers in the relays, the proposed ODWF scheme could better exploit the time
diversity of microscopic fading.

\subsection{Relaying Schemes}

The proposed ODWF relay scheme with fixed relays and microscopic
fading channels is given below.

\begin{Scheme}[ODWF Scheme for Fixed Relays]
The buffer of each relay is divided into $N$ equal-sized banks,
each for the packets received on one subcarrier. The scheduling
protocol is described below:
\begin{itemize}
\item[1.] Each relay measures the states of its links with the
source and the destination in the channel estimation slot.

\item[2.] The control slot is divided into two halves, called
control sub-slots 1 and 2:

\begin{itemize}
\item If relays can forward packets to the destination on each
subcarrier, a random backoff mechanism is performed in the control
sub-slot 1 and one relay per subcarrier is selected to forward the
packet in the transmission slot.

\item Otherwise, if there are connected source-relay links on each
subcarrier, the relays will ask the BS for new packets in sub-slot
2.
\end{itemize}

\item[3.] If some relays are authorized to deliver packets in the
transmission slot, these relays will transmit the first packet in
the queue for the selected subcarrier. In the meantime, all other
relays should listen and delete from their the buffers the same
packet as the transmitted packets. On the other hand, if the
source is authorized to deliver packets in the transmission slot,
it will transmit new packets on all $N$ subcarriers. Otherwise, no
transmission happens in the transmission slot.
\end{itemize}\label{sch:micro-op}
\end{Scheme}

Furthermore, the existing relay design is also presented as a
baseline for evaluating the performance of the proposed. The baseline scheme is
a regular decode-and-forward scheme without buffers.

\begin{Baseline}[Regular Decode-and-Forward Scheme for Fixed Relays]
\
\begin{itemize}
\item[1.] Each relay measures the states of its links with the
source and the destination in the channel estimation slot.

\item[2.] The control slot is divided into two halves, called
control sub-slots 1 and 2:

\begin{itemize}
\item If there is no packet in the relay network, the relays will
ask the BS for new packets in sub-slot 1 (The frame is used for
phase I).

\item If there are packets in the relay network, a cooperative
mechanism is performed in sub-slot 2 to determine which relay will
transmit which packet on which subcarrier in the transmission
slots\footnote{In this paper, we shall compare with the
performance upper bound of the baseline. Therefore, we assume the
cooperative mechanism is genie-aided.}(The frame is used for Phase
II).
\end{itemize}

\item[3.] If some relays are selected in the control slot, these
relays will forward packets on the selected subcarrier to the
destination in the transmission slot. Otherwise, if the source is
selected in the control slot, it will transmit new packets on all
$N$ subcarriers in the transmission slot.
\end{itemize}
\label{sch:micro-al}
\end{Baseline}

\subsection{Throughput-Delay Tradeoff of The ODWF Scheme with Fixed
Relays}

With the opportunistic decode-wait-and-forward scheme, when the
source increases the packet data rate, the number of relays that
can decode the packets from the source becomes small and then, the
probability that these packets can be forwarded to the destination
in the next frame also becomes small due to the less multi-relay
diversity, leading to a large end-to-end average packet delay.
Therefore,
there is a tradeoff between the average system throughput $T$ and
the average end-to-end packet delay $D$ when ODWF is used in the
system. This
tradeoff is summerized in the following theorem.

\begin{Theorem}[Throughput-Delay Tradeoff with Fixed Relays] Denote
the data
rate as $r = \log_2\left(1+p\ln
\beta\right)$. Suppose Scheme \ref{sch:micro-op} is implemented,
for infinite buffers at relays and sufficiently large $K$,

\begin{itemize}
\item[I.] The achievable average system throughput is
upper-bounded by
\begin{equation}
T_{max} \doteq \frac{N}{2}\log_2
\left[1+p\ln(K)\right],\label{eqn:avg_tp}
\end{equation}
and this throughput is achievable under the following conditions:
\begin{equation}
\lim\limits_{K\rightarrow +\infty} \frac{\ln \beta}{\ln K}=1
\quad\mbox { and }\quad \lim\limits_{K\rightarrow +\infty}
\frac{\beta}{K} = 0. \nonumber
\end{equation}

\item[II.] If $\lim\limits_{K\rightarrow +\infty} \frac{\beta}{K}
= 0$, then the average system throughput is given by
\begin{equation}
T \doteq \frac{N}{2}\log_2 \left(1+p\ln\beta\right),
\label{eqn:mi-tp}
\end{equation}
and the average end-to-end packet delay is given by
\begin{equation}
D \doteq \max\left\{1,\frac{2c\beta^2}{K}\right\}  , \quad where
\quad c = \ln\left(\frac{2^{1/N}}{2^{1/N}-1}\right).
\label{eqn:mi-delay}
\end{equation}
\end{itemize}
\label{the:infty-relay}
\end{Theorem}

\begin{proof}
Please refer to Appendix A.
\end{proof}

The throughput-delay tradeoff in the above theorem can be
interpreted as follows. The threshold $\beta$ separates the
multi-relay diversity for enhancing throughput and for reducing
delay. A larger $\beta$ corresponds to higher throughput but
longer end-to-end packet delay and vice versa.
\begin{itemize}
\item \textbf{Case I - Small Data Rate ($\lim\limits_{K\rightarrow
+\infty} \frac{\beta}{\sqrt{K}} = 0$)}. Suppose the source
transmits $N$ packets on $N$ subcarriers respectively in one
frame, for any subcarrier there are sufficient number of relays
that can decode the source's packet (connected links). So in the
next frame, because of sufficient multi-relay diversity, for any
subcarrier there are some relays (with large probability) that can
forward the received packet to the destination. Hence, the source
and the relay transmissions will occur alternately, and the Scheme
\ref{sch:micro-op} will reduce to the Scheme \ref{sch:micro-al}.

\item \textbf{Case II - Large Data Rate
($\lim\limits_{K\rightarrow +\infty} \frac{\beta}{\sqrt{K}} > 0$
and $\lim\limits_{K\rightarrow +\infty} \frac{\beta}{K} = 0$)}
Suppose the source transmits $N$ packets on $N$ subcarriers
respectively in one frame, the number of relays that can decode
the packets is small for each subcarrier. Hence, it can not be
guaranteed that these packets can be forwarded to the destination
in the next frame due to the lack of multi-relay diversity. As a
result, the packets are stored in the relays' buffers until
channels becomes reliable. In other words, when the spatial
diversity is insufficient, the ODWF scheme has to further exploit
the time diversity, increasing the end-to-end packet delay.
Therefore, the
larger the data rate, the less the relays who can decode the
packets, and then, the larger the end-to-end packet delay. Moreover,
it can
be deduced from the above theorem that when the maximum average
system throughput is achieved, the corresponding average end-to-end
packet
delay is $\mathcal{O}(K)$.
\end{itemize}

Notice that the multi-relay diversity contributes to the average
system throughput with the order of $\ln\ln \beta$ according to
(\ref{eqn:mi-tp}), which actually grows very slowly with respect
to $K$ (as $\beta\sim\mathcal{O}(K)$); on the other hand, the
resultant end-to-end packet delay increases much faster with
$\beta$, namely $\beta^2$  as shown in (\ref{eqn:mi-delay}). The
multi-relay diversity in the fixed relay network with microscopic
fading channels does not improve the system throughput
significantly. Nevertheless, we shall show in Section
\ref{sec:rlt-macro} that exploiting the multi-relay diversity in
the mobile relay networks would bring much significant throughput
gains.

\subsection{Comparison with The Regular Decode-and-Forward Scheme
(Baseline
\ref{sch:micro-al})}

Providing the throughput-delay tradeoff derived in the preceding
part, we could continue to show the performance gain of the
proposed ODWF scheme (Scheme \ref{sch:micro-op}) over that of the
baseline scheme (Baseline \ref{sch:micro-al}). We first introduce
the following lemma:

\begin{Lemma}[Performance of Regular Decode-and-Forward Scheme with
Fixed
Relays] The maximum average throughput of the Baseline
\ref{sch:micro-al} with $K$ relays is given by
\begin{equation}
T \doteq \frac{N}{2}\log_2 \left(1+p\ln \sqrt{K}\right),
\label{eqn:mi-tp-al}
\end{equation}
while the average end-to-end packet delay $D\doteq 1$ \label{cor:fix}
\end{Lemma}

\begin{proof}
Please refer to Appendix B.
\end{proof}

Comparing (\ref{eqn:mi-tp}) and (\ref{eqn:mi-tp-al}), we have the
following corollary:

\begin{Corollary}
The Scheme \ref{sch:micro-op} with $\sqrt{K}$ relays and a cost of
$\mathcal{O}(K)$ average end-to-end packet delay can achieve the
same optimal average throughput as the Baseline \ref{sch:micro-al}
with $K$ relays.
\end{Corollary}

\begin{figure}
 \begin{center}
  \resizebox{15cm}{!}{\includegraphics{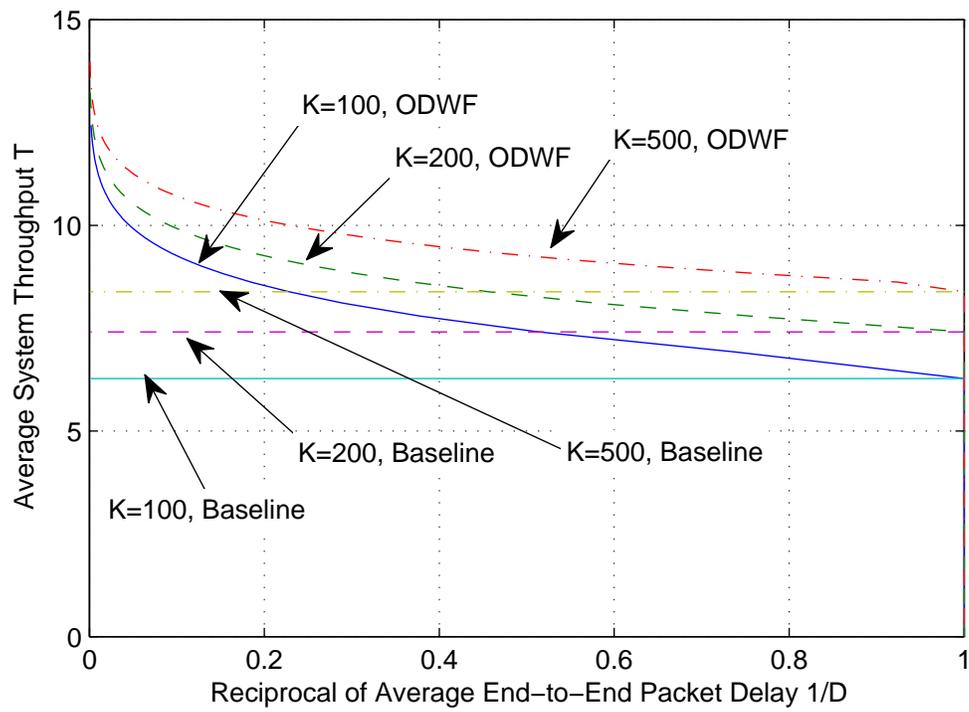}}
 \end{center}
    \caption{Throughput-delay tradeoff in the fixed relay scenario.}
    \label{fig:fixsim}
\end{figure}

This corollary can be interpreted as follows. If Scheme
\ref{sch:micro-op} is implemented, the source just need to
guarantee the best $\Theta(1)$ links on each subcarrier are
connected links and hence, $\beta=\mathcal{O}(K)$. On the other
hand, if Baseline \ref{sch:micro-al} is implemented, one packet
can be forwarded to the destination when at least
$\Theta(\sqrt{K})$ relays can decode it. Therefore, the source
should guarantee the best $\Theta(\sqrt{K})$ links on each
subcarrier are connected links, leading to $\beta =
\mathcal{O}(\sqrt{K})$. Fig. \ref{fig:fixsim} illustrates the
throughput-delay tradeoffs of the ODWF scheme and the baseline
scheme for fixed relays according to Theorem \ref{the:infty-relay}
and Lemma \ref{lem:micro-tp}. To normalize the x-axis, we plot the
reciprocal of the average end-to-end delay $\frac{1}{D}$ instead.
As it's shown in this figure, the ODWF scheme has the same
throughput-delay tradeoff as the baseline scheme in small delay
($D=1$), and achieves larger throughput at the expense of larger
delay.

\section{Opportunistic decode-wait-and-forward Scheme for Mobile
Relays with Macroscopic
Fading Channels}
\label{sec:rlt-macro}

In this section, we first describe the proposed opportunistic
decode-wait-and-forward relay scheme for mobile relays and
macroscopic fading channels, and then discuss its performance. We show that by
utilizing buffers in the relays, we could better exploit the mobility of relays
and achieve a significantly better performance compared with regular decode and
forward relay protocols.

\subsection{Relaying Schemes}

The proposed ODWF scheme for mobile relays and macroscopic fading
channels is described below:

\begin{Scheme}[ODWF Scheme for Mobile Relays]
\
\begin{itemize}
\item[1.] Each relay measures the pathloss to the source and the
destination at the channel estimation slot.

\item[2.] The control slot is divided into two halves, called
control sub-slots 1 and 2. The following is the system procedure
operating in the control slot:

\begin{itemize}
\item If there exist relays with packets in their buffers and
connected link to the destination, a random backoff mechanism is
operated in sub-slot 1 to determine which relay could access the
destination in the transmission slot.

\item Otherwise, if there exist relays with connected links to the
source, these relays will notify the source in sub-slot 2 to
transmit a new packet in the transmission slot.
\end{itemize}

\item[4.] If one relay is picked up for the transmission slot,
this relay will transmit the first packet in its buffer. In the
meantime, all other relays should listen and delete from their
buffers the same packet. On the other hand, if the source is
picked up for the transmission slot, it will broadcast one new
packet. Otherwise, no transmission happens in the transmission
slots.
\end{itemize}\label{sch:macro-op}
\end{Scheme}

In order to demonstrate the performance gain of the ODWF scheme,
the existing relay scheme is also presented in the following as a
baseline of comparison.

\begin{Baseline}[Regular Decode-and-Forward Scheme for Mobile Relays]
\
\begin{itemize}
\item[1.] Each relay measures the path loss of its links with the
source and the destination in the channel estimation slot.

\item[2.] The control slot is divided into two halves, called
control sub-slots 1 and 2:

\begin{itemize}
\item If there is no packet in the relay network, the relays will
ask the BS for one new packet in sub-slot 1 (The frame is used for
phase I).

\item If there is one packet in the relay network, a cooperative
mechanism is performed in sub-slot 2 to determine which relay will
forward this packet in the transmission slot (The frame is used
for Phase II).
\end{itemize}

\item[3.] If one relay is selected in the control slot, this relay
will forward the packet to the destination in the transmission
slot. Otherwise, if the source is selected in the control slot, it
will transmit one new packet to the relay network in the
transmission slot.
\end{itemize}
\label{sch:macro-al}
\end{Baseline}

\subsection{Throughput-Delay Tradeoff of ODWF Scheme with Mobile
Relays}

Since the transmit power is constant, when the source increases
the data rate, its radio coverage become small and the number of
relays who can decode this packet also becomes small. Hence, the
probability these relays can forward the decoded packet to the
destination is small, leading to a large end-to-end packet delay.
Therefore,
there is a tradeoff between the system throughput and the end-to-end
packet
delay, which is presented rigorously in the following theorem.

\begin{Theorem}[Throughput-Delay Tradeoff with Mobile Relays] Denote
the data
rate as $r=N\log_2 \beta$.
Suppose the Scheme \ref{sch:macro-op} is implemented, for infinite
buffers at relays, sufficiently large $K$, and positive $q$:
\begin{itemize}
\item[I.] The achievable average system throughput of the relay
system is upper-bounded by
\begin{equation}
T_{max} \doteq \frac{N\alpha}{4} \log_2 K. \label{eqn:ma-max-tp}
\end{equation}
where $\doteq$ denotes the equality with high probability (\whp),
and this throughput is achievable under the following conditions:
\begin{equation}
\lim\limits_{K\rightarrow +\infty} \frac{\ln
\beta^{\frac{2}{\alpha}}}{\ln K}=1 \quad\mbox { and }\quad
\lim\limits_{K\rightarrow +\infty}
\frac{\beta^{\frac{2}{\alpha}}}{K} = 0. \nonumber
\end{equation}

\item[II.] If $\lim\limits_{K\rightarrow +\infty}
\frac{\beta^{\frac{2}{\alpha}}}{K}=0$, the average system throughput
is given by
\begin{equation}
T \doteq \frac{N}{2}\log_2 \beta \label{eqn:ma-tp} ,
\end{equation}
and the average end-to-end packet delay is given by
\begin{equation}
D \doteq
\mathcal{O}\left(\max\{\frac{\beta^{\frac{4}{\alpha}}}{Kq},
\frac{1}{q}\}\right).\label{eqn:ma-delay}
\end{equation}
\end{itemize}\label{the:macro-tradeoff}
\end{Theorem}

\begin{proof}
Please refer to Appendix C.
\end{proof}

\textit{Remarks:} (I) Since there are infinitely large buffers at
relays and the region transition probability $q$ is positive, the
average system throughput is $r/2$ as long as there is always
relays having connected links to the source and destination (which
is presented mathematically as $\lim\limits_{K\rightarrow +\infty}
\frac{\beta^{\frac{2}{\alpha}}}{K}=0$).

(II) There are two factors affecting the average end-to-end delay
expression (\ref{eqn:ma-delay}), namely $\beta^{\frac{4}{\alpha}}$
and $\frac{1}{q}$. The first factor is related to the packet data
rate $r$. When $r$ is large (so as the $\beta$), the source's
coverage becomes small and hence, the number of relays who can
decode the packet from the source is small. Therefore, this packet
should wait for many frames until these relays carry it into the
coverage of the destination. Due to the random walk of the relays'
movement, the less the relays who can decode this packet, the
longer time on average this packet should wait. On the other hand,
the second factor is relate to the velocity of the relays'
movement. If the velocity is small (so as the region transition
probability $q$), statistically one relay should use long time
before moving into the coverage of the destination, which also
leads to large end-to-end packet delay.

(III) Compared with the throughput of $\mathcal{O}(\ln \ln K)$ in
the scenario of fixed relays, exploiting the relays' mobility
could bring the system much larger throughput. Noting that the
frame duration is usually milliseconds for the fixed relay
scenario but seconds (or even larger) for the mobile relay
scenario, the large throughput is obtained at the expense of
delay.

\begin{figure}
 \begin{center}
  \resizebox{15cm}{!}{\includegraphics{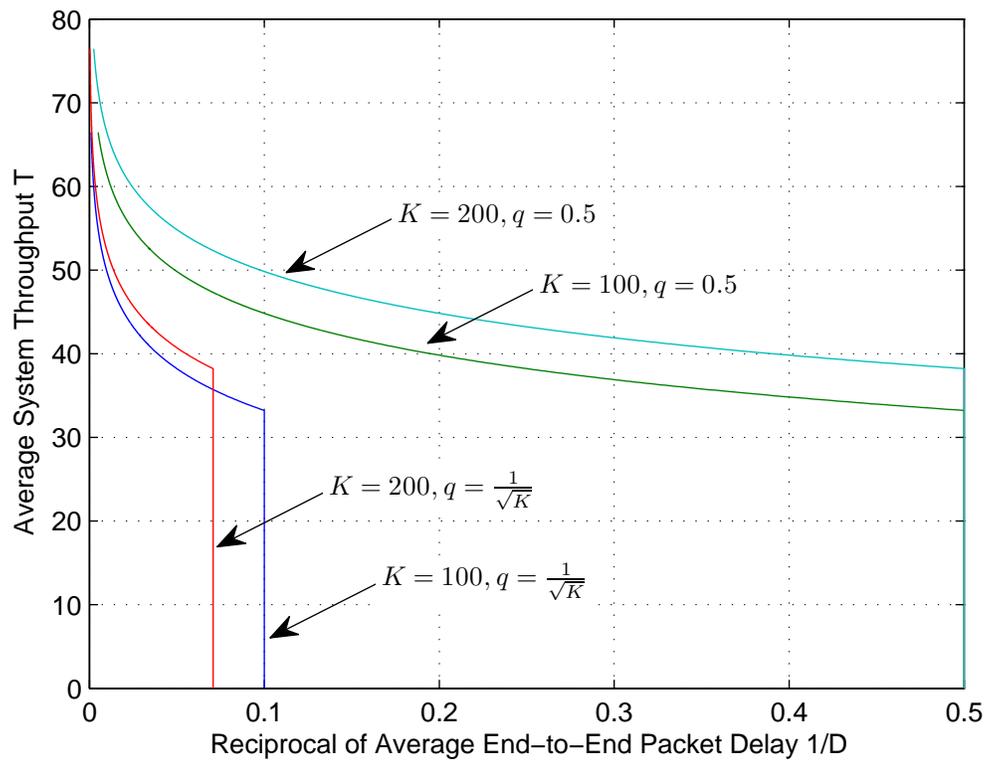}}
 \end{center}
    \caption{Throughput-delay tradeoff in the mobile relay scenario.}
    \label{fig:mobilesim}
\end{figure}

According to Theorem \ref{the:macro-tradeoff}, we plot the
throughput-delay tradeoff of the ODWF scheme for mobile relays in
Fig. \ref{fig:mobilesim}. To normalize the x-axis, the reciprocal
of the average end-to-end delay $\frac{1}{D}$ is plotted instead.
As it's shown in this figure, larger $K$ would lead to larger
average system throughput since the system can enjoy more
multi-relay diversity. Furthermore, smaller relays' velocity $q$
would lead to larger average end-to-end packet delay for a given
throughput level, having the curve shrink towards the y-axis.

\subsection{Comparison with Regular Decode-and-Forward Scheme
(Baseline
\ref{sch:macro-al})}

Before the comparison is made, the performance of the baseline is
summarized in the
following lemma:

\begin{Lemma}[Performance of Regular Decode-and-Forward Scheme with
Mobile
Relays] For non-zero region-transition-probability $q$ and
sufficiently large number of relays $K$, the following statements
on the Baseline \ref{sch:macro-al} are true:
\begin{itemize}
\item[I.] If
$\lim\limits_{K\rightarrow+\infty}qK^{\frac{1}{M-1}}<+\infty$, the
maximum achievable average system throughput is given by
\begin{equation}
T_{max} \doteq \mathbf{\Theta}\left(qK^{\frac{1}{M-1}}\right),
\end{equation}
and the corresponding average end-to-end packet delay is $D\doteq
\mathbf{\Theta}(\frac{1}{Kq^{M-1}})$.

\item[II.] If
$\lim\limits_{K\rightarrow+\infty}qK^{\frac{1}{M-1}}=\infty$, the
maximum achievable average system throughput is given by
\begin{equation}
T_{max} \doteq \mathbf{\Theta}\left(\log_2 K\right),
\end{equation}
and the corresponding average end-to-end packet delay is $D\doteq
\mathbf{\Theta}(1)$.

\end{itemize}\label{lem:macro-base}
\end{Lemma}

\begin{proof}
Please refer to the Appendix D.
\end{proof}

Therefore, we have the following corollary on the performance gain
of the ODWF scheme:

\begin{Corollary}
Let $T_o$ and $T_r$ be the optimal average system throughput of
the Scheme \ref{sch:macro-op} and Baseline \ref{sch:macro-al},
then when
$\lim\limits_{K\rightarrow+\infty}qK^{\frac{1}{M-1}}=\infty$,
\begin{equation}
\frac{T_o}{T_r} = \mathbf{\Theta}(1);\nonumber
\end{equation}
and when
$\lim\limits_{K\rightarrow+\infty}qK^{\frac{1}{M-1}}<\infty$,
\begin{equation}
\frac{T_o}{T_r} = \mathbf{\Theta}\left(\frac{\log_2
K}{qK^{\frac{1}{M-1}}}\right).\nonumber
\end{equation}
\end{Corollary}

Although when the relays' movement is fast
($\lim\limits_{K\rightarrow+\infty}qK^{\frac{1}{M-1}}=\infty$) the
average system throughput of the ODWF scheme and the baseline is
at the same order, the ODWF could still obtain a gain of constant
scaling. This is because the proposed ODWF scheme can exploit the
"idle frame" intelligently before the relays with packets moving
into the destination's coverage. The slower the relays' movement,
the more such "idle frames" and hence, the larger the throughput
gain of the ODWF scheme. Therefore, a orderwise gain can be
obtained by ODWF scheme when the relays' movement is slow, (e.g.
$q=\frac{1}{K}$).

\section{Conclusions} \label{sec:con}

In this paper, we propose an opportunistic decode-wait-and-forward
(ODWF) scheme for a point-to-point communication system with $K$
relays. The throughput-delay tradeoff of this ODWF scheme is
studied in two scenarios: fixed relays and mobile relays, and the
performance gain over the existing relay schemes in the same
scenarios is also examined. In the scenario of fixed relays, the
ODWF scheme could exploit the spatial and temporal dynamics of
multipath fading channels so that it can achieve the same optimal
average throughput as the existing schemes with $K^2$ relays. In
the scenario of mobile relays, the ODWF scheme could exploit the
dynamic locations of relays so that the average system throughput
can scale as $\Theta(\ln K)$, which can not be achieved by the
existing schemes when the relay's mobility is slow. Moreover,
compared with the well-known throughput scaling law of the
amplify-and-forward relay networks, where the $\Theta(\log K)$
throughput is achieved with $\Theta(K)$ total transmission power,
we show that by exploiting the relays' mobility the same order of
throughput can be achieved even with constant transmission power.

\section*{Appendix A: Proof of Theorem \ref{the:infty-relay}}

We first introduce and prove the following lemma:

\begin{Lemma}
When $\lim\limits_{K\rightarrow +\infty}
\frac{\beta}{\sqrt{K}}=+\infty$ and the system has been runing for
sufficiently long time, on each subcarrier there are $\alpha
\times 100\%$ percentage of relays containing packets in their
buffer with high probability, where
\begin{equation}
\alpha =
\frac{\ln\left[1-\frac{\delta}{(1+\delta^N)^{\frac{1}{N}}}\right]}{
\ln(1-\delta)
}\nonumber
\end{equation}
and $\delta = 1-(1-\frac{1}{\beta})^K$. Moreover, the average
system throughput is given by
\begin{equation}
T \doteq \frac{N\delta^N}{1+\delta^N} \log_2[1+p\ln
\beta].\nonumber
\end{equation}\label{lem:micro-tp}
\end{Lemma}

\begin{proof}
When the source transmits packets to the relay cluster,  on each
subcarrier there are approximately $e^{-\beta}K$ relays can decode
the packet. Hence, the probability that the next frame is used for
relay-destination transmission is $
\left[1-(1-\frac{1}{\beta})^{\frac{K}{\beta}}\right]^N$. Since
$\lim\limits_{K\rightarrow +\infty} \frac{\sqrt{K}}{\beta}=0$, we
have
\begin{equation}
\lim_{K\rightarrow +\infty}
\left[1-(1-\frac{1}{\beta})^{\frac{K}{\beta}}\right]^N = 0.
\nonumber
\end{equation}
Thus, those received packets could hardly be forwarded to the
destination immediately in the next frame. Instead, they will be
stored in the relays' buffers temporarily. As new packets
transmitted from the source become more and more, on every
subcarrier the ratio of relays who have packets stored in their
buffers become larger and larger. The increase of the ratio will
continue until the following flow balance equation is satisfied
\begin{equation}
P_{RD} = P_{SR},\label{eqn:flow_bal}
\end{equation}
where $P_{SR}$ is  the probability one frame is used for
source-relay transmission and $P_{RD}$ is the probability one
frame is used for relay-destination transmission. Assume the
corresponding ratio is $\alpha
* 100 \%$ \textit{whp}. Since
\begin{equation}
P_{RD}=\left[1-(1-\frac{1}{\beta})^{\alpha K}\right]^N \nonumber
\end{equation}
and
\begin{eqnarray}
P_{SR} &=& (1-P_{RD})[1-(1-\frac{1}{\beta})^{K}]^N, \nonumber
\end{eqnarray}
according to(\ref{eqn:flow_bal}), we can obtain that
\begin{equation}
\alpha =
\frac{\ln\left[1-\frac{\delta}{(1+\delta^N)^{\frac{1}{N}}}\right]}{
\ln(1-\delta)
}\nonumber
\end{equation}
and
\begin{equation}
P_{SR}=P_{RD} = \frac{\delta^N}{1+\delta^N},\label{eqn:p_rd}
\end{equation}
where $\delta = 1-(1-\frac{1}{\beta})^K$. As a result, the average
throughput is given by
\begin{eqnarray}
T & \doteq & Nr \times P_{RD} \nonumber\\
& \doteq & \frac{N\delta^N}{1+\delta^N}
\log_2[1+p\ln\beta].\nonumber
\end{eqnarray}
where the notation $\doteq$ is necessary due to the randomness of
the channel fading.

\end{proof}

\subsection{The proof of statement I:}

According to Lemma \ref{lem:micro-tp}, the achievability is
verified below:
\begin{eqnarray}
T &\doteq&\frac{N\delta^N}{1+\delta^N} \log_2[1+p\ln\beta]
\nonumber\\
&\rightarrow&\frac{N}{2}\log_2[1+p\ln\beta] \label{eqn:st-1}\\
&=&\frac{N}{2}\log_2[1+p\ln K] \label{eqn:st-2} ,
\end{eqnarray}
where (\ref{eqn:st-1}) is due to the condition
$\lim\limits_{K\rightarrow +\infty}(1-\frac{1}{\beta})^K=0$, and
(\ref{eqn:st-2}) is due to $\lim\limits_{K\rightarrow +\infty}
\frac{\ln \beta}{\ln K} = 1$. In the following, we shall prove
that there is no threshold which could lead to a larger average
throughput.

It's easy to see that when $\lim\limits_{K\rightarrow +\infty}
\frac{\ln\beta}{\ln K} < 1$, the average system throughput $T$ is
less than $\frac{N}{2}\log_2[1+p\ln K]$. When
$\lim\limits_{K\rightarrow +\infty} \frac{\ln\beta}{\ln K}
> 1$
\begin{eqnarray}
T&\doteq&\frac{N\delta^N}{1+\delta^N} \log_2[1+p\ln\beta]\nonumber\\
&=&N\delta^N
\log_2[1+p\ln\beta]\nonumber\\
&=& N\left[1-(1-\frac{1}{\beta})^K\right]^N
\log_2[1+p\ln\beta]\nonumber\\
&\rightarrow&0 \quad (as \ K\rightarrow +\infty)\nonumber .
\end{eqnarray}

\subsection{The proof of the statement II}

The proof the this statement is divided into two parts. We shall
prove the case when $\lim_{K\rightarrow +\infty}
\frac{\beta}{\sqrt{K}}=0$. If the source transmits packets to the
relay cluster, on each subcarrier there are approximately
$\frac{K}{\beta}$ relays can decode the packet. Hence, the
probability that the next frame is used for relay-destination
transmission is $
\left[1-(1-\frac{K}{\beta})^{\frac{K}{\beta}}\right]^N$. Notice
that when $\lim_{K\rightarrow +\infty} \frac{\beta}{\sqrt{K}}=0$,
\begin{equation}
\lim_{K\rightarrow +\infty}
\left[1-(1-\frac{K}{\beta})^{\frac{K}{\beta}}\right]^N = 1.
\nonumber
\end{equation}
The relay-destination transmission will follow the source-relay
transmission with probability $1$, thus, the phase I and phase II
will occur alternately. Therefore, the average end-to-end packet
delay $D$ is $1$ \whp  and the average system throughput $T$ is
$\frac{r}{2}=\frac{N}{2} \log_2[1+p\ln\beta]$ \whp.

Now we turn to the case when $\lim_{K\rightarrow +\infty}
\frac{\beta}{\sqrt{K}}>0$ and $\lim_{K\rightarrow +\infty}
\frac{\beta}{K}\leq 1$. The derivation of the average system
throughput $T$ can follow the same approach as we did in the proof
of statement I. Hence, we only show how to derive the average
end-to-end packet delay $D$ in the following.

Suppose the system has been running for sufficiently long time,
according to Lemma \ref{lem:micro-tp}, there are $\alpha \times
100\%$ percent of relays on each subcarrier having packets in the
buffers. Notice that $\lim\limits_{K\rightarrow \infty} \alpha=0$,
when the source delivers one new packet on each subcarrier, there
are $\frac{K}{\beta}$ relays who can decode the packet on each
subcarrier, and almost all of them have empty buffer for this
subcarrier. Hence, the new arrival packets are on the head of
relays' queues. Notice that $\frac{K}{\beta}<<\alpha K$ and the
relay selection if phase II is uniformly random, the probability
that one new arrival packet is forward by the relay cluster is
given by $P_{RD}\frac{1}{\alpha \beta}$, where $P_{RD}$ is the
probability one frame is used for relay-destination transmission
which is given by (\ref{eqn:p_rd}). As a result, the average
end-to-end packet delay can be written as
\begin{eqnarray}
D &\doteq& \frac{\alpha \beta}{P_{RD}}\nonumber\\
& = &2\alpha\beta\nonumber\\
&=&2\beta\frac{\beta}{K}\ln\left(\frac{2^{1/N}}{2^{1/N}-1}
\right)\nonumber\\
&=&\frac{2c\beta^2}{K} \quad where \quad c =
\ln\left(\frac{2^{1/N}}{2^{1/N}-1}\right)\nonumber ,
\end{eqnarray}
where the second equality is because $P_{RD}\approx \frac{1}{2}$;
the third equality is because $\alpha \rightarrow
\frac{\beta}{K}\ln\left(\frac{2^{1/N}}{2^{1/N}-1}\right)$ for
$K\rightarrow +\infty$ according to Lemma \ref{lem:micro-tp}.
Combine the results in the case of $\lim_{K\rightarrow +\infty}
\frac{\beta}{\sqrt{K}}=0$, (\ref{eqn:mi-delay}) is proved.

\section*{Appendix B: The Proof of Lemma \ref{cor:fix}}

The achievability is verified below. Let $\beta=\sqrt{K}/\ln K$.
When the source transmits packets to the relay cluster, there are
$\frac{K}{\beta}=\sqrt{K} \ln K$ relays on each subcarrier who can
decode the packets. Hence, on every subcarrier, the chance the
received packets can be forwarded to the destination in the next
frame is $1-\left(1-\frac{1}{\beta}\right)^{\sqrt{K} \ln
K}\rightarrow 1$. In other words, packets in the relay cluster can
be forward to the destination in the next frame \whp, and the
average throughput is given by:
\begin{eqnarray}
T&\doteq&\frac{N}{2}\log_2\left[1+p(\ln\sqrt{K} -\ln \ln
K)\right]\nonumber\\
&\approx&\frac{N}{2}\log_2(1+p\ln \sqrt{K}). \label{eqn:alt_tp}
\end{eqnarray}

In the following, we shall continue to show that there is no
$\beta$ which can lead to an average throughput larger than
$\frac{N}{2}\log_2(1+p\ln \sqrt{K})$. Thus, three cases are
discussed:
\begin{itemize}
\item $\lim\limits_{K\rightarrow +\infty} \frac{\ln\beta}{\ln
K}\leq \frac{1}{2}$: It's obvious that the average throughput in
this case can not be larger than $\frac{N}{2}\log_2(1+p\ln
\sqrt{K})$.

\item $\frac{1}{2}<\lim\limits_{K\rightarrow +\infty}
\frac{\ln\beta}{\ln K}\leq 1$: After the source's transmission,
there are $\frac{K}{\beta}$ relays on each subcarrier who can
decode the packets. Hence, for any packet in the relays' buffers,
it can choose one connected link out of $NK/\beta$ links.
Therefore, the probability this packet can be delivered to the
destination is upper bounded by
$1-\left(1-\frac{1}{\beta}\right)^{NK/\beta}$, and the average
throughput is upper bounded by
$\left[1-\left(1-\frac{1}{\beta}\right)^{NK/\beta}\right]\frac{r}{2}
\rightarrow
0$ (when $K\rightarrow +\infty$).

\item $\lim\limits_{K\rightarrow +\infty} \frac{\ln\beta}{\ln
K}>1$: It's with high probability that no relay can decode the
packets from the source in this case. Hence, the throughput of
this case tends to zero.

\end{itemize}

In conclusion, the average throughput in (\ref{eqn:alt_tp}) is
maximum for Scheme \ref{sch:micro-al}.

\section*{Appendix C: The Proof of Theorem \ref{the:macro-tradeoff}}

We first prove the second statement, from which the first
statement can be deduced.

\subsection*{The Proof of The Statement II - Average System
Throughput}

\begin{Lemma}
For the threshold $\beta$ satisfying $\lim\limits_{K\rightarrow
+\infty} \frac{\beta^{\frac{2}{\alpha}}}{K}=0$ and sufficiently
large $K$, the probability that there are
$\mathbf{\Theta}(\frac{K}{\beta^{\frac{2}{\alpha}}})$ relays
having connected links with the source (or the destination) tends
to $1$.
\end{Lemma}

\begin{proof}
Without loss of generality, we only study the number of connected
links with the source. The connection with the destination can
follow the same approach. It's equivalent to examine the another
relay distribution where the relays' location on the disk is
uniform and i.i.d. between frames.

Since we are interested in the analysis with the sufficiently
large $K$ and the monotonically increasing data rate $r$
($\lim\limits_{K\rightarrow +\infty} r=+\infty$), the coverage
radii of the source and the destination in our analysis are
sufficiently small given the fixed transmit power $p$. Therefore,
for relays with connected links to the source/destination, the
link capacity defined in (\ref{eqn:cp-sj}) and (\ref{eqn:cp-jd})
can be approximated as
\begin{equation}
C^{s,j}\approx N\log_2\frac{p}{d_{s,j}^{\alpha}} \quad \mbox{and}
\quad C^{j,d} \approx N\log_2\frac{p}{d_{j,d}^{\alpha}},\nonumber
\end{equation}
and the coverage radius of the source within which the relays have
connected links to the source is given by
\begin{equation}
d=\frac{p^{\frac{1}{\alpha}}}{\beta^{\frac{1}{\alpha}}}.\nonumber
\end{equation}
Therefore, the area of the source's coverage can be approximated
as $\frac{\pi d^2}{2}$, and the probability one relay falls into
the source's coverage is $\frac{d^2}{2R^2}$.

Suppose there are $X$ relays having connected links with the
source, therefore,
\begin{eqnarray}
\Pr(X=x)&=&  \binom{K}{x}
\left(\frac{d^2}{R^2}\right)^x\left(1-\frac{d^2}{R^2}\right)^{K-x}.
\end{eqnarray}
From the above equation, the probability that $X =
\mathbf{\Theta}(\frac{d^2}{R^2}K)$ tends to 1 when $K\rightarrow
+\infty$.
\end{proof}

Notice that there are (with probability $1$) some relays in region
$1$ can decode the source's packets, the relay buffer is infinite,
relays are moving among the divided regions, and there are (with
probability $1$) some relays in region $M$ can forward packets to
the destination, the average system throughput is
\begin{equation}
T \doteq \frac{r}{2} = \frac{N}{2}\log_2\beta.\nonumber
\end{equation}

\subsection*{The Proof of The Statement II - Average Packet Delay}

We study the system in the following two cases: (1)
$\lim\limits_{K\rightarrow
+\infty}\frac{\beta^{\frac{4}{\alpha}}}{K}=0$, and
(2)$\lim\limits_{K\rightarrow
+\infty}\frac{\beta^{\frac{4}{\alpha}}}{K}>0$ and
$\lim\limits_{K\rightarrow
+\infty}\frac{\beta^{\frac{2}{\alpha}}}{K}=0$. The average
end-to-end packet delay for the these two cases is given by the
following two lemmas.

\begin{Lemma}
For the threshold $\beta$ satisfying $\lim\limits_{K\rightarrow
+\infty}\frac{\beta^{\frac{4}{\alpha}}}{K}=0$, the average
end-to-end packet delay is given by $D=\mathcal{O}(\frac{1}{q})$
\end{Lemma}

\begin{proof}
When the source broadcasts a packet, there are
$\mathbf{\Theta}\left(\frac{K}{\beta^{\frac{2}{\alpha}}}\right)$
relays can decode this packet. Since the region transition
probability is $q$, after $\mathbf{\Theta}(1/q)$ frames, there are
$\mathbf{\Theta}\left(\frac{K}{\beta^{\frac{2}{\alpha}}}\right)$
relays with this packet moves into the region $M$. Then, this
packet can be forwarded to the destination with probability $1$,
because
\begin{equation}
\frac{K}{\beta^{\frac{2}{\alpha}}}
\frac{1}{\beta^{\frac{2}{\alpha}}} \rightarrow +\infty \mbox{ when
} K\rightarrow +\infty.
\end{equation}
\end{proof}

\begin{Lemma}
For the threshold $\beta$ satisfying $\lim\limits_{K\rightarrow
+\infty}\frac{\beta^{\frac{4}{\alpha}}}{K}>0$ and
$\lim\limits_{K\rightarrow
+\infty}\frac{\beta^{\frac{2}{\alpha}}}{K}=0$, the average
end-to-end packet delay is given by
$D=\mathcal{O}(\frac{\beta^{\frac{4}{\alpha}}}{Kq})$
\end{Lemma}

\begin{proof}
To prove the average packet delay for the second case is
$\mathcal{O}(\frac{\beta^{\frac{4}{\alpha}}}{Kq})$, we first suppose
\begin{equation}
\lim_{K\rightarrow+\infty}\frac{D}{\frac{\beta^{\frac{4}{\alpha}}}{Kq
}}=\infty\nonumber,
\end{equation}
and then show the contradiction in the following.

Let $D=\frac{\beta^{\frac{4}{\alpha}}}{Kq}\delta$, where
$\lim\limits_{K\rightarrow +\infty} \delta = +\infty$. The average
packet delay is $D$ implies that the average number of packets
stored in the relay network is $\mathbf{\Theta}(D)$. Hence, there
are $\mathbf{\Theta}(\frac{\beta^{\frac{4}{\alpha}}}{Kq}\delta)$
packets stored in the relay networks for at least
$\mathbf{\Theta}(1/q)$ frames. Therefore, it can be easily seen
that there are $\mathbf{\Theta}(\beta^{\frac{2}{\alpha}}\delta)$
relays with packets in region $M$, which implies that the relays
will deliver packets to the destination in every frame. Hence,
this assumption is impossible.
\end{proof}

\subsection*{The Proof of The Statement I}

The achievability can be verified by letting
$\beta=\left(\frac{K}{\ln^2 K}\right)^{\frac{\tau}{2}}$. In order
to prove the throughput (\ref{eqn:ma-max-tp}) is maximum,two cases
are discussed in the following:

\begin{itemize}
\item Case I: $\lim\limits_{K\rightarrow
+\infty}\frac{\beta}{K^{\frac{\tau}{2}}} =C$ where $C$ is a
constant. It's easy to verify that in this case the average system
throughput can not be larger than (\ref{eqn:ma-max-tp}).

\item Case II: $\lim\limits_{K\rightarrow
+\infty}\frac{\beta}{K^{\frac{\tau}{2}}} =+\infty$. The
probability that there is no relay in the source's coverage is
$\left[1-(\frac{p}{\beta})^{\frac{2}{\tau}}\frac{M}{2R^2}\right]^{
\frac{K}{M}}$,
hence the average system throughput is given by
\begin{eqnarray}
T&=&\left\{1-\left[1-(\frac{p}{\beta})^{\frac{2}{\tau}}\frac{M}{2R^2}
\right]^{
\frac{K}{M}}\right\}\frac{N}{2}\log_2
\beta\nonumber\\
&\approx&(\frac{p}{\beta})^{\frac{2}{\tau}}\frac{K}{2R^2}\frac{N}{2}
\log_2
\beta \nonumber\\
&\rightarrow& 0  \ \ (as \ \ K\rightarrow +\infty \nonumber).
\end{eqnarray}
\end{itemize}

This completes the proof of Theorem \ref{the:macro-tradeoff}.

\section*{Appendix D: The Proof of Lemma \ref{lem:macro-base}}

When the source transmit one packet to the relay network, the
probability that there are
$\mathbf{\Theta}\left(\frac{K}{\beta^{\frac{2}{\alpha}}}\right)$
relays can decode the packet tends to $1$ (following the same
approach as in Appendix C). Then, after $t$ frames, the number of
relays who moves into the region 2 from the region 1 is
$\mathbf{\Theta}\left(\frac{K}{\beta^{\frac{2}{\alpha}}}qt\right)$;
the number of relays who moves into the region 3 from the region 2
from is
$\mathbf{\Theta}\left(\frac{K}{\beta^{\frac{2}{\alpha}}}(qt)^2\right)$,...,
the number of relays who moves into the region M-1 from the region
M from is
$\mathbf{\Theta}\left(\frac{K}{\beta^{\frac{2}{\alpha}}}(qt)^{M-1}\right)$.
In order to deliver the packet to the destination, the average
end-to-end packet delay $D$ should satisfy
\begin{equation}
\frac{K}{\beta^{\frac{2}{\alpha}}}(qD)^{M-1} \doteq
\mathbf{\Theta}(\beta^{\frac{2}{\alpha}}).\nonumber
\end{equation}
Therefore, it can be derived from the above equation that
\begin{equation}
D \doteq \mathbf{\Theta} \left[ \max\left\{\frac{1}{q}
\left(\frac{\beta^{\frac{4}{\alpha}}}{K}\right)^{\frac{1}{M-1}} ,
1\right\}\right],\nonumber
\end{equation}
and the average system throughput is given by
\begin{eqnarray}
T &\doteq & \frac{N}{2D}\log_2 \beta\nonumber\\
&=&\mathbf{\Theta} \left[ \min\left\{\frac{Nq}{2}
\left(\frac{K}{\beta^{\frac{4}{\alpha}}}\right)^{\frac{1}{M-1}}\log_2
\beta , \frac{N}{2}\log_2 \beta \right\}\right].\nonumber
\end{eqnarray}

\subsection*{The Proof of The Statement I}

When $\lim\limits_{K\rightarrow +\infty}q
K^{\frac{1}{M-1}}<+\infty$, the average system throughput becomes
\begin{equation}
T \doteq \mathbf{\Theta}\left(\frac{Nq}{2}
\left(\frac{K}{\beta^{\frac{4}{\alpha}}}\right)^{\frac{1}{M-1}}\log_2
\beta\right). \nonumber
\end{equation}
After maximizing with respect to $\beta$
($\beta=\mathbf{\Theta}(1)$, we have
\begin{equation}
T_{max} \doteq \mathbf{\Theta}\left(\frac{Nq}{2}
K^{\frac{1}{M-1}}\right) = \mathbf{\Theta}\left(q
K^{\frac{1}{M-1}}\right).\nonumber
\end{equation}

\subsection*{The Proof of The Statement II}

When $\lim\limits_{K\rightarrow +\infty}q
K^{\frac{1}{M-1}}=+\infty$, the  optimal $\beta$ which maximizes
the average system throughput is
\begin{equation}
\beta^{\frac{4}{\alpha}} =  \mathbf{\Theta} (q K^{\frac{1}{M-1}}).
\nonumber
\end{equation}
Therefore,
\begin{equation}
T \doteq \mathbf{\Theta}\left(\frac{N}{2} \log_2
q^{\frac{\alpha}{4}} K^{\frac{\alpha}{4(M-1)}}\right) =
\mathbf{\Theta}\left(\log_2 K\right). \nonumber
\end{equation}
This finishes the proof of Lemma \ref{lem:macro-base}.

\bibliographystyle{IEEEtran}
\bibliography{Relay_Mobility_Ver5.2}

\newpage

\end{document}